\documentclass[11pt]{article}

\usepackage[utf8]{inputenc}
\usepackage{amsmath} 
\usepackage{color}
\usepackage{multirow}
\usepackage[unicode]{hyperref}
\usepackage{lineno}
\usepackage{geometry}  % Flexible and complete interface to document dimensions

\usepackage{lineno}  % line numbers on paragraphs
\usepackage{fancyhdr}  % Extensive control of page headers and footers in LaTeX2ε
\usepackage{enumitem}  % Control layout of itemize, enumerate, description
\usepackage{hyperref}  % Extensive support for hypertext in LaTeX
\usepackage{titling}  % Control over the typesetting of the \maketitle command
\usepackage[numbers,compress]{natbib}  % Flexible bibliography support
\usepackage{filecontents}
\usepackage{multirow}
\usepackage{mathtools}  % Mathematical tools to use with amsmath
\usepackage{titlesec}  % Select alternative section titles
\usepackage{color}
\usepackage{lastpage}  % Reference last page for Page N of M type footers
\usepackage{subcaption}
\usepackage[english]{babel}  % Multilingual support for Plain TeX or LaTeX
\usepackage{amsmath}  % AMS mathematical facilities for LaTeX
\usepackage{amsfonts}  % TeX fonts from the American Mathematical Society
\usepackage{textgreek}
\usepackage{amssymb}  % Additional symbols from American Mathematical Society
\usepackage{wasysym}  % LaTeX support file to use the WASY2 fonts
\usepackage{bbm}  % "Blackboard-style" cm fonts
\usepackage{array}  % Extending the array and tabular environments
\usepackage{xr}  % References to other LaTeX documents
\usepackage{verbatim} % Reimplementation of and extensions to LaTeX verbatim
\usepackage{float} % Improved interface for floating objects
\usepackage{marginnote}
\usepackage{todonotes}
\usepackage{ulem}

\usepackage{siunitx}
\newcommand{\SIadj}[2]{\SI[number-unit-product={\text{-}}]{#1}{#2}}

\hypersetup{%
  pdfborderstyle={/S/U/W 1}% border style will be underline of width 1pt
}

\titleformat{name=\section}{\normalfont\Large\bfseries}{}{0pt}{}
\titleformat{name=\subsection}{\normalfont\large\bfseries}{}{0pt}{}
\titleformat{name=\subsubsection}{\normalfont\normalsize\bfseries}{}{0pt}{}

\newcommand{\email}[1]{\href{mailto:{#1}}{{#1}}}

\newcommand{\keywords}[1]{\textbf{Keywords}: {#1}}

\newcommand{\optincludegraphics}[2][]{}
\newcommand{\optinput}[1]{}
\newtagform{brackets}{[}{]}
\usetagform{brackets}

\newcommand{\thejournal}[1]{Nature Communication}

\title{{A Self-Matched Leaky-Wave Antenna\newline for Ultrahigh-Field MRI with Low SAR}}

\usepackage{ifthen}
\usepackage{suffix}
\newcommand\figref[2][]{Figure~\figref*[#1]{#2}}
\newcommand\figrefp[2][]{Figures~\figref*[#1]{#2}}
\WithSuffix\newcommand\figref*[2][]{\ref{#2}\ifthenelse{\equal{#1}{}}{}{#1}}

\newcommand{\abs}[1]{\ensuremath{\left|#1\right|}}

\newcommand{\wmag}{\ensuremath{w_\text{mag}}}
\newcommand{\wel}{\ensuremath{w_\text{el}}}

\begin{document}

\begin{titlepage}
{\noindent\LARGE\bf \thetitle}

\bigskip

\begin{flushleft}\large
	G. Solomakha\textsuperscript{1},
    J. T. Svejda\textsuperscript{2},
	C. van Leeuwen\textsuperscript{3},
	 A. Rennings\textsuperscript{2},
    A. J. Raaijmakers\textsuperscript{{3},4},
    S. Glybovski\textsuperscript{{1},{*}},
    D. Erni\textsuperscript{2},

\end{flushleft}

\bigskip

\noindent
\begin{enumerate}[label=\textbf{\arabic*}]
\item Faculty of Physics and Engineering, ITMO University, St. Petersburg, Russia
\item General and Theoretical Electrical Engineering (ATE), Faculty of Engineering, University of Duisburg-Essen, and CENIDE -- Center for Nanointegration Duisburg-Essen, 47048 Duisburg Germany
\item Imaging Division, UMC Utrecht, Utrecht, the Netherlands
\item 
Medical Image Analysis, Biomedical Engineering, Technical University of Eindhoven, Eindhoven, The Netherlands

\end{enumerate}

\bigskip

\textbf{*} Corresponding author:
\indent\indent
\begin{tabular}{>{\bfseries}rl}
Name		& Stanislav Glybovski										\\
Department	& Faculty of Physics and Engineering		\\
Institute	& ITMO University												\\
Address 	& 49 Kronverksky Pr.											\\
			& 197101														\\
            & Russian Federation														\\
E-mail		& \email{s.glybovski@metalab.ifmo.ru}							\\
\end{tabular}

\vfill

Running Head: Leaky-wave coil with low SAR
%\vfill
%Manuscript words: 5389

\end{titlepage}

\pagebreak

\begin{abstract}
The technology of magnetic resonance imaging is developing towards higher magnetic fields to improve resolution and contrast.
However, whole-body imaging at \num{7} T or even higher fields remains challenging due to wave interference, tissue inhomogneities and high RF power deposition.
Nowadays, proper RF excitation of a human body in prostate and cardiac MRI is only possible to achieve by using phased arrays of antennas attached to the body (so-called surface coils).
Due to safety concerns, the design of such coils aims to minimize the local specific absorption rate (SAR) keeping the highest possible RF signal in the region of interest.
All previously demonstrated approaches were based on resonant structures such as e.\ g.\ dipoles, capacitively-loaded loops, TEM-line sections.
In this study, we show that there is a better compromise between the transmit signal \(B_1^{+}\) and the local SAR using non-resonant surface coils due to weaker RF near fields in the close proximity of their conductors.
With this aim, we propose and experimentally demonstrate a first leaky-wave surface coil implemented as a periodically-slotted microstrip transmission line.
Due to its non-resonant radiation, the proposed coil induces only slightly over half the peak local SAR compared to a state-of-the-art dipole coil, but has the same transmit efficiency in prostate imaging at \SI{7}{\tesla}.
Unlike other coils, the leaky-wave coil intrinsically matches its input impedance to the averaged wave impedance of body tissues in a broad frequency range, which makes it very attractive for future clinical applications of \SI{7}{\tesla} MRI.

\bigskip
\end{abstract}

\pagebreak

Currently there is a clear trend towards higher static magnetic fields (\(B_0\)) in magnetic resonance imaging (MRI) systems. 
The interest in so-called ultrahigh-field (UHF) scanners, having \(B_0\) of \SI{7}{\tesla} and higher for imaging and spectroscopy of human body tissues, is explained by higher achievable image resolution and contrast with shorter examination times as compared to the widespread \SI{1.5}{\tesla} and \SI{3}{\tesla} clinical systems~\cite{ladd2018}.
Thanks to the availability of \SI{7}{\tesla} whole-body scanners, it is possible to increase our knowledge of biology and medicine by collecting more data, e.\ g.\ in the functional investigation of a human brain~\cite{TRATTNIG2018477} or of a spinal cord~\cite{BARRY2018437}.
The \SI{7}{\tesla} technology even provides insight into mechanisms of neuropsychiatric disorders with much better reliability~\cite{Morris2019}.
However, despite the recent advances in biomedical research using UHF MRI, this technology is still not implemented in clinics. 
The main factors preventing clinicians from using UHF MRI for medical diagnostics originate from undesired interference effects of propagating electromagnetic waves in body tissues at relatively high Larmor frequencies. 
Thus for protons at \SI{7}{\tesla} the Larmor frequency increases to \SI{298}{\mega\Hz}, so that the average wavelength in body tissues shrinks to \SI{13}{\centi\m}~\cite{Collins_IMA}, which is comparable to e.\ g.\ the extent of internal organs in the abdominal cavity~\cite{7Tesla_Body}.
Along with high attenuation of electromagnetic waves due to the relatively high conductivity of the medium, interference effects make the radio frequency (RF) magnetic field intrinsically inhomogeneous~\cite{Cloos2016}. 
As a result, body images at 7T inevitably have dark voids~\cite{7Tesla_Body}.
 
The inhomogeneity issue is typically addressed in research UHF MR systems by the method of parallel transmission (pTx)~\cite{pTx}.
Unlike birdcage coils used for whole-body imaging at the low frequencies of clinical scanners~\cite{birdcage}, pTx allows for the manipulation of the transmit field distribution.
This is achieved by using multiple surface coils placed directly onto a body and driven with customized phases and amplitudes, which allows to steer the signal voids away from the region of interest.
This approach, however, is not allowed in clinical MRI as it typically requires careful preliminary determination of individual transmit phases and constancy of coil tuning and matching for each subject.
Another limitation is the potentially high peak local specific absorption rate (SAR) created by each antenna element typically due to its close proximity to the body surface.
Peak SAR strongly depends on the geometry of the coil as it is related to electric fields in the near field region and needs to be minimized.
At the same time, properly designed transmit surface coils should maximize the signal in the region of interest (ROI) for a given applied power.
The signal is proportional to the correlating components of the circularly polarized RF magnetic field \(B_1^{+}\).
In the case of abdominal cavity imaging at \SI{7}{\tesla} (in particular, prostate imaging), the ROI is located one or more wavelengths away from the surface coil and can be considered as an intermediate field region.
In this region, the electromagnetic field already resembles a propagating wave rather than a quasi-static field~\cite{SSAD}.
Therefore, on-body coils creating weaker electric fields at a body surface and stronger magnetic fields in the center of a body are preferable especially in prostate and cardiac UHF imaging due to higher transmit efficiency and improved safety. 

Transmit surface coils have traditionally consisted of surface loops~\cite{pTx} arranged parallel to the body surface. 
In the low-frequency approximation, such loops are efficient at small penetration depths, where their quasi-static magnetic field, normally polarized with respect to the surface, is rather strong. 
However, since the far field of a vertical magnetic dipole in its axial direction is zero, surface loops become inefficient. 
Indeed, at \SI{7}{\tesla} \(B_1^{+}\)-magnetic field produced by a surface loop is strongly inhomogeneous. 
Moreover, to reach deeply embedded ROIs the dimensions of a surface loop must be large, which makes them self-resonant and further distorts their field pattern~\cite{NMR_probeheads}. 
For this reason, alternative design types of surface coils have been proposed. 
Stripline resonators~\cite{Striplines} allow a more densely packed transmit array without using decoupling circuits. 
Higher Q-factors of stripline resonators lead to higher currents in the coil and hence, to higher \(B_1^{+}\) for the same transmit power, but only for the quasi-static near field region. 
In~\cite{SSAD} it was shown that deeply located imaging targets in \SI{7}{\tesla} whole-body imaging are located outside the near field region. 
To better image such ROIs coils must be designed to redistribute the magnetic RF energy towards the intermediate- or even far-field region (using \emph{radiative} coils~\cite{Alex_11}). 
One of the most efficient types of radiative surface coils used for transceiver body arrays at \SI{7}{\tesla} is the dipole~\cite{Alex_11}. 
Among several known configurations of dipoles used e.\ g.\ for prostate imaging at \SI{7}{\tesla}, \emph{fractionated} dipoles~\cite{SSAD}  have demonstrated the best compromise between the transmit efficiency for deeply located targets and the peak local SAR~\cite{Alex_MRM}. 
A dipole oriented along the \(B_0\) field typically creates the maximum RF magnetic field for the given depth of the ROI right under its center regardless of distance. 
Like a dipole antenna in free space, it operates efficiently when its length becomes comparable to a half-wavelength. 
In other words, though a dipole has a much smaller Q-factor compared to a loop, a standing wave of current still occurs along its conductors. 
The corresponding \(B_1^{+}\)-field induces an almost sinusoidal half-wave pattern in the body. 
Conveniently, the radiation of dipoles placed over a high-permittivity half-space is dominant towards the dielectric medium~\cite{Engheta_dipoles}. 
In MRI this allows dipoles to efficiently convert their radiated power into dissipative losses inside the body.

Recently, it was demonstrated that two parallel dipoles with in-phase currents combined into the same coil further reduce the peak local SAR as compared to a single dipole due to weaker near fields and lower Q~\cite{Solomakha_2019_MRM}. 
As alternatives to dipoles and loops, other surface coil types have been shown to be suitable for operation at \SI{7}{\tesla}, such as transverse slots~\cite{Alon_MRM} and even water-filled slotted waveguide resonators~\cite{Ilic}. 

Noticeably, all previously designed surface coils known from the literature are resonant.
The only known non-resonant excitation method, called \emph{traveling-wave MRI}, used the cylindrical bore (RF shield of a scanner) as a waveguide in which, at Larmor frequencies larger than \SI{270}{\mega\Hz}, a propagating \(\text{TE}_{11}\) mode is supported. 
This mode, typically excited with a patch antenna, delivers the RF signal to a human body or a head~\cite{Brunner_TW, HERRMANN201801, Anna_diel}.
Despite providing low SAR, this method strongly suffers from low transmit efficiency and weak parallel transmit capabilities. 
As for efficient surface body coils for pTx, one can find a direct correspondence between the Q-factor of the resonator used in a coil and the peak local SAR induced for the same \(B_{1}^{+}\) at large depths in a body.

\begin{figure*}[tbp]
  \centering
  \includegraphics[width=1.0\textwidth]{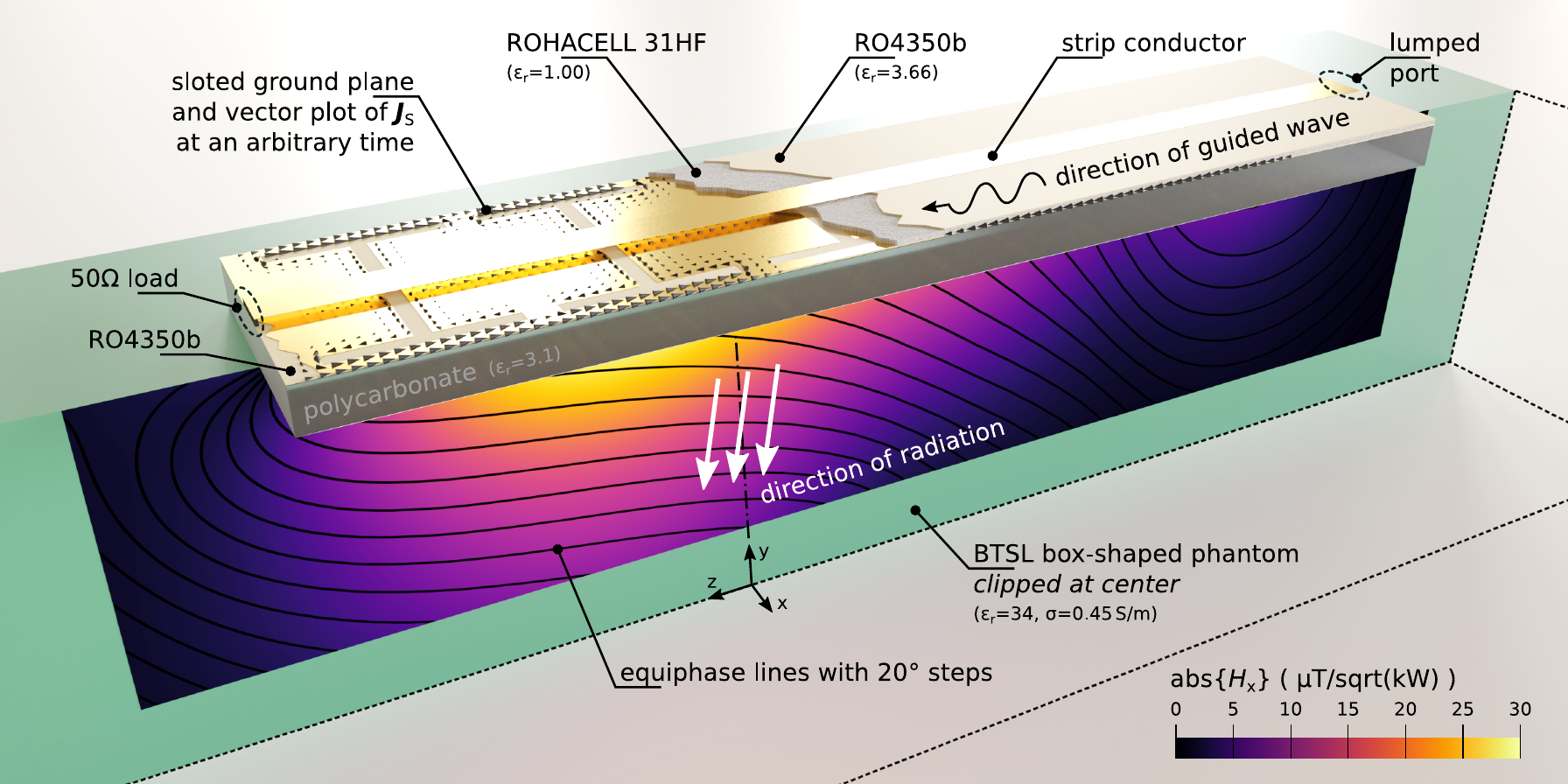}
  \caption{%
    RF-excitation of an MRI phantom with the proposed leaky-wave coil (LWC).
    The colour plot shows the normalised magnitude of the simulated \(H_{x}\) distribution, while black contours represent wavefronts in the conductive medium.
    Calculated current distribution on the slotted ground plane is shown with black arrows.%
    }%
  \label{fig:View}
\end{figure*}%
In this study, we propose a new excitation structure for UHF body imaging based on a leaky-wave antenna (LWA) element.
We use a non-resonant surface coil based on a carefully-designed slotted microstrip line that launches leaky waves into a human body when placed on it as schematically depicted in \figref{fig:View}.
The proposed coil induces a much lower SAR for the same \(B_1^{+}\) efficiency when imaging deeply located regions of the human body (i.\ e.\ prostate, kidneys) as compared to state-of-the-art RF-coils.
Moreover, due to the leaky-wave radiation employed, its input impedance is intrinsically broad-band matched to the transmitter.%

\section{Results}
\subsection{Resonant and Leaky-wave RF excitation in MRI}

To our knowledge, all previously proposed surface coils, such as e.\ g.\ dipoles, loops and stripline segments, consist of resonators placed close to a human body. 
In other words, their operation is based on the excitation of standing waves. 
This approach necessarily results in the excitation of strong reactive electric and magnetic fields in the vicinity of a coil. 
It is well-known in antenna engineering that the higher the Q-factor and reactance, the stronger the reactive near fields.
In UHF body imaging, the magnetic components in the reactive near field are concentrated near the surface coil and do not contribute to the signal in deeply located ROIs (e.\ g.\ in the prostate).
Whereas its electric field components cause local SAR hotspots. 

\begin{figure*}[tbp]
  \centering
  \includegraphics[width=0.85\linewidth]{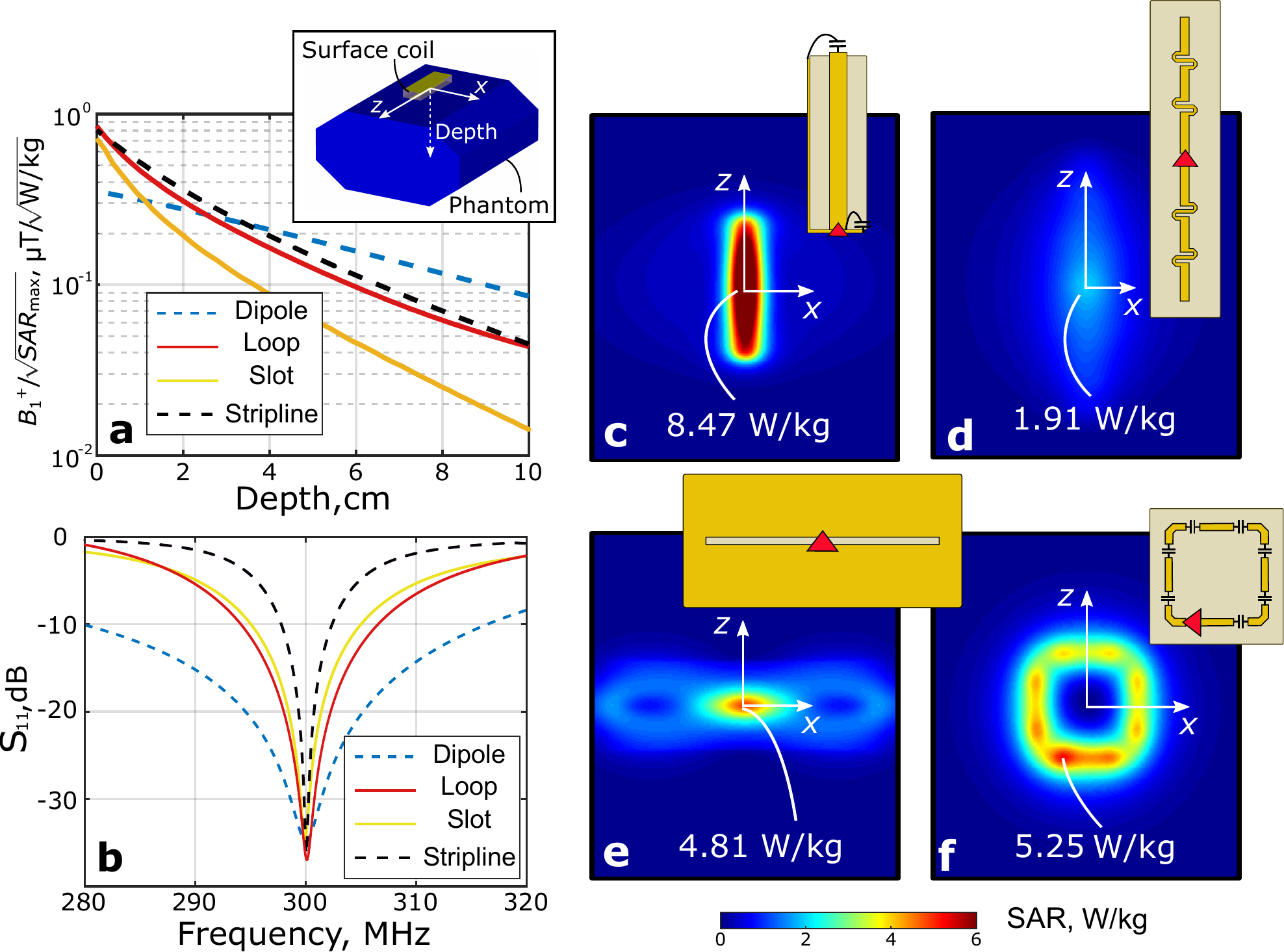}
  \caption{%
    Numerical proof of the correspondence between the peak local SAR and resonant properties of four different surface coil types used for body imaging at \SI{7}{\tesla}: (a) $B_{1}^{+}/\sqrt{\textnormal{SAR}}$ factor for different depths of ROI; (b) reflection coefficient at the input of a coil; simulated local SAR patterns at \SI{1}{\W} of accepted transmit power at the top surface of the phantom: (c) stripline; (d) fractionated dipole; (e) slot; (f) loop.
  }%
  \label{fig:SAR}
\end{figure*}%
To illustrate the relationship between the peak local SAR and the quality factor, we have numerically analyzed four of the most popular surface coils used as elements of body transmit arrays at \SI{7}{\tesla}.
In the simulation, the coils were placed on top of a pelvis-shaped body phantom (\(\varepsilon_\text{r} = \num{34}\) and \(\sigma = \SI{0.45}{\siemens\per\m}\)~\cite{Alex_MRM}) with a \SIadj{1}{\centi\m}-thick spacer.
All four coils were tuned at \SI{300}{\mega\Hz} and matched to \SI{50}{\ohm} with appropriate \(\pi \)-circuits of lumped elements and accept the same power of \SI{1}{\W} from the transmitter.
\figref{fig:SAR} shows the results of the comparison.
From signal profiles normalized by the square root of peak SAR (the so-called \emph{SAR efficiency}) in \figref[a]{fig:SAR} and from SAR profiles in \figref[c-f]{fig:SAR}, it follows that the compared coil types are very different in terms of RF safety (i.\ e.\ they create different SAR levels for the same of accepted power).
\figref[b]{fig:SAR} shows the frequency dependencies of the reflection coefficient \(S_{11}\) at the feed port of each coil, which illustrates the bandwidth of the impedance being inversely proportional to the Q-factor.
The stripline coil created the highest SAR for the lowest \(B_{1}^{+}\) at depths larger than \SI{5}{\centi\m}, while its bandwidth was the smallest.
For the fractionated dipole the result is the opposite.
Comparing the maximal SAR values and bandwidths of the coils, one can clearly check that the broader the bandwidth of the coil, the lower the peak SAR it creates at the phantom surface (in its near field region).

\begin{figure}
  \centering
  \includegraphics[width=0.4\textwidth]{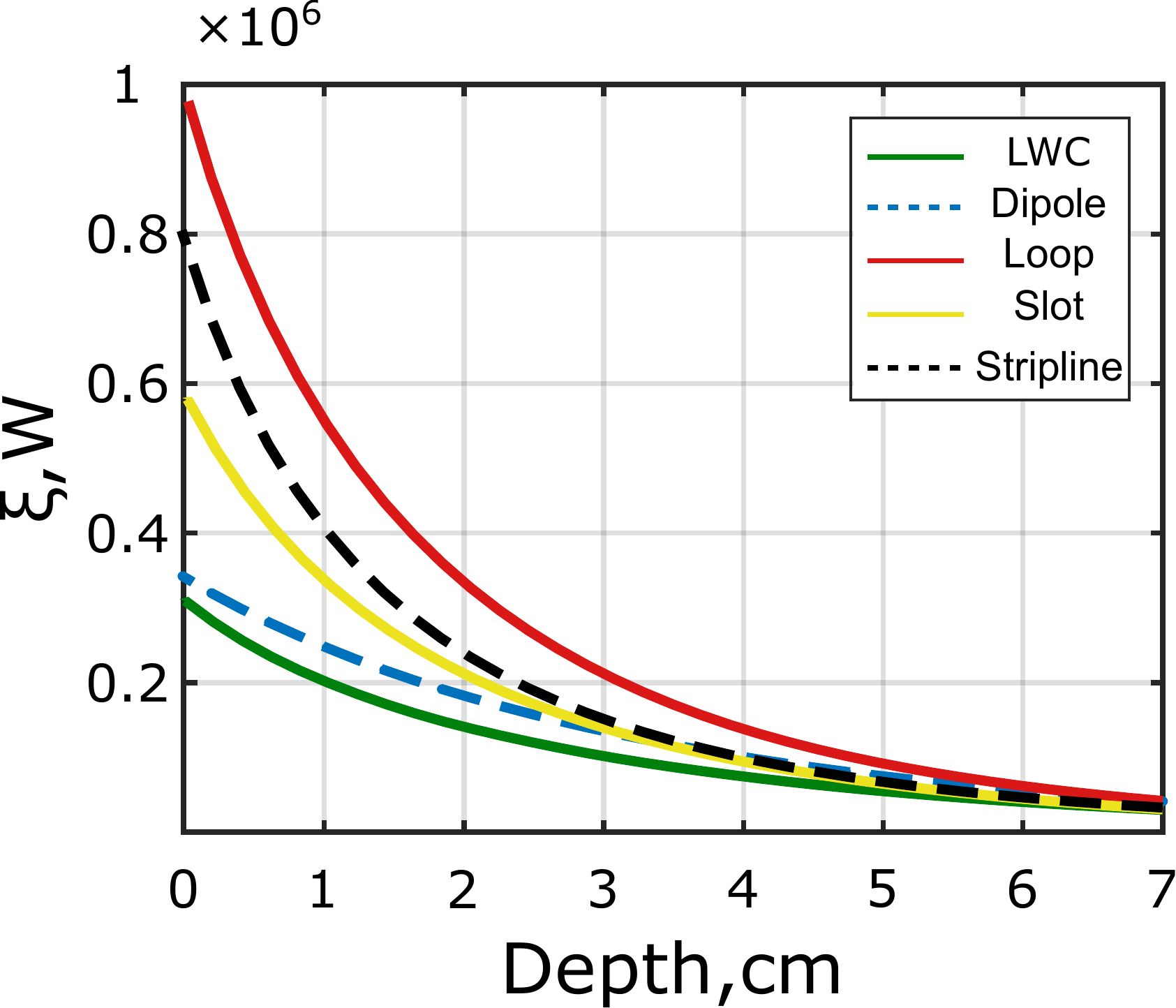}
  \caption{%
  Difference of energy density based figure of merit defined in equation~\eqref{eqn:integral} for different coil types equally fed with \SI{1}{\W} of accepted power as a function of depth in a phantom.
  }%
  \label{fig:EnergyDiffernceFourCoils}
\end{figure}%

This relation can be explained using Poynting's theorem written in the differential form with respect to the region of the phantom. Within this region electromagnetic sources are absent and only ohmic  losses can be considered. In this case, for the time dependence taken in the form $\exp(j\omega t)$, the Poynting's theorem reads
\begin{align}
  \nabla \cdot \vec{S} &= -\frac{1}{2}\sigma\abs{\vec{E}}^2 - \mathrm{j}\omega\frac{1}{2}\left( \mu\abs{\vec{H}}^2 - \varepsilon_0\varepsilon_{\text{r}}\abs{\vec{E}}^2\right) \nonumber \\
    &= -\frac{1}{2}\sigma\abs{\vec{E}}^2 - \mathrm{j}2\omega  \Delta w,
    \label{eqn:generalDifferenceOfEnergyDesity}
\end{align}
where \(\vec{S}\) is the Poynting vector, and \(\Delta w\) is the difference of time-averaged magnetic and electric energy densities \(\Delta w = \wmag - \wel = \frac{1}{4}\mu\abs{\vec{H}}^2 - \frac{1}{4}\varepsilon_0\varepsilon_{\text{r}}\abs{\vec{E}}^2\).
\(\Delta w\) vanishes in lossless regions permeated by propagating wave fields, whereas a significant difference occurs in regions of dominant reactive near fields, typically around resonating structures, due to the imposed spatial separation of electric and magnetic energy densities.

The spatial distribution of \(\Delta w\) can thus be used as a qualitative measure to visualize the distinction between the domain of a coil's reactive near field and the adjacent region where (attenuated) propagating field solutions are assumed to emerge. 

In lossy media, represented here by the permittivity and the conductivity of the human body tissue, the difference \(\Delta w\) remains finite even for the latter case of a propagating wave because of induced electric currents and their additional magnetic field contribution.
This attenuated wave addresses therefore the least resonant state in the lossy medium yielding the smallest possible difference \(\Delta w\), which is thus associated to a minimal reactive field.
By considering equation~\eqref{eqn:generalDifferenceOfEnergyDesity}, large values of \(\Delta w\) are implicitly related to considerable conductive losses via involved electric fields respective electric currents.
Large, spatially separated but resonant electric and magnetic energy densities are interlinked by balancing current flows that are likely to contribute to enhanced SAR values.

Based on equation~\eqref{eqn:generalDifferenceOfEnergyDesity} one can now introduce a comprehensive measure \(\xi\) for the presence of reactive fields created by the different coil types inside the phantom.
Given the definition
\begin{align}
    \xi(y) &= \int\limits_{V_{\text{int}}(y)} 2\omega \abs{\Delta w} \mathrm{d}V 
    \label{eqn:integral}
\end{align}
where the absolute value of \(\Delta w\) is integrated over the \(y\)-dependent volume \(V_\text{int}\), which covers the whole phantom from the outermost boundary at \(y = d_\text{y}\) up to a given horizontal cross-section at position \(y\).
Here, \(d_\text{y}-y\) is labeling the volume's variable extent between this horizontal cross-section and the far boundary at \(d_\text{y}\) of the entire phantom.
Therefore, \(\xi(y)\) is perfectly apt to visualize the spatial behavior of the (horizontally averaged) reactive near-fields around the coil in relation to the best case of an attenuated propagating wave field as a function of the penetration depth \(y\) into the phantom, which is shown in \figref{fig:EnergyDiffernceFourCoils} for all four coil types.
As can be seen from these graphs, the loop coil shows the largest reactive fields but is not considered the worst coil in terms of the resulting SAR (cf.\ \figref[a]{fig:SAR}).
This demonstrates that \(\xi\) or even \(\Delta w\) cannot always be directly related to SAR due to the importance of the involved field distributions.
Nevertheless, from \figref{fig:EnergyDiffernceFourCoils} it is evident that the proposed LWC creates the lowest amounts of reactive fields as compared to the other coils and hence excites mainly propagating waves as desired.

To design a surface coil with even lower reactive field and SAR having $B_{1}^{+}$ the same as a fractionated dipole, we proposed to use a non-resonant structure operating as a leaky-wave artificial transmission line.
The idea of the proposed structure is illustrated in \figref{fig:View} and it is similar to the radiation of leaky-wave antennas (LWAs), typically operating in the microwave range and above.
In such antennas, a propagating wave along the length of the line continuously radiates into free space.
According to~\cite{LWA_jackson}, one of the main types of LWAs is called the \emph{uniform LWA}, in which the guiding structure is uniform along its length and supports a wave that is fast with respect to the speed of light in free space.
In other words, the phase constant \(\beta \) in the complex propagation constant \(k_\text{z} = \beta - \mathrm{j}\alpha \) stays in the range \(0 < \beta < k_0\), where \(k_0\) is the free-space wavenumber.
In this case, the waveguide structure is periodic (an artificial transmission line with the possibility to adjust the wave dispersion~\cite{lai2004composite}) and the periodicity is small compared to the wavelength, so the LWA is classified as a \emph{quasi-uniform LWA}.
LWAs have been popular due to their low-profile implementation and the possibility to steer a narrow radiation beam by tuning the frequency.
Usually such antennas are based on slotted metallic waveguides and periodically loaded (defected) microstrip lines.
The phase constant \(\beta \) of the guided wave is adjusted to define the radiation angle \(\theta_\text{rad} \) through the approximate relation \(\sin\theta_\text{rad} = \beta/k_0\) at given frequencies within the whole hemisphere~\cite{caloz2008crlh}.
At the same time, the attenuation factor (leakage constant) \(\alpha \) is chosen to maximize the gain by radiating about \SI{90}{\percent} of the applied power at the propagation length \(L\)~\cite{6174425}.
As long as the wave loses most of its power in the form of radiation during its propagation from the feeding port to the matched load at the end of the line, there is no resonant behavior, hence the input impedance of such antennas is relatively stable with respect to frequency.

LWAs operate in free space, and therefore, should support a fast propagating wave to radiate efficiently.
In contrast, surface coils for MRI should operate when positioned in close proximity to a human body, which represents a highly conductive medium with high permittivity.
In this case, leaky-wave radiation is allowed even if \(\beta \) is comparable to \(k_0\), i.\ e.\ for simple TEM-lines. Indeed, as it has been previously demonstrated, even a simple slot line, which cannot radiate when situated in free space, becomes an efficient and ultrawideband leaky-wave radiator when placed at a boundary of a half-space with high permittivity~\cite{bruni2007ultrawideband}.
This effect has been used to achieve radiation into high-permittivity media in such applications as microwave lens antennas~\cite{Neto} and ground penetrating radar~\cite{Deiana}.
When speaking about radiation beneath the boundaries of conductive media, leaky-wave applicators have been demonstrated as appropriate sources of inhomogeneous (exponentially decaying) waves which can cause the effect of \emph{deep electromagnetic penetration} into a conductive medium~\cite{baccarelli2018analytical,taylor1984penetrating}. 
Along with the non-resonant properties of leaky-wave radiation possibly causing lower near fields, the possibility to use the deep penetration effect to reach high transmit efficiency in deeply located ROIs was our motivation to study the operation of a leaky-wave surface coil for MRI.

To reach the leaky-wave excitation at the Larmor frequency of protons at 7 T (around \SI{300}{\mega\Hz}), we designed our coil as a microstrip-line section with a periodically slotted ground plane. 
The phantom considered here has the averaged properties of human abdominal tissues, i.\ e.\ relative permittivity \(\varepsilon_\text{r,ph} = \num{34}\) and conductivity \(\sigma_\text{ph} = \SI{0.4}{\siemens\per\m}\). 
In this case, the guided wave in quasi-uniform microstrip line is fast with respect to a plane wave in the medium of the phantom and leaky-wave radiation becomes possible.
The line with a strip width of \SI{15}{\milli\m} and height of \SI{2}{\milli\m} is matched to the transmitter at its input port. 
The ground plane of the microstrip is separated from the strip by a \SIadj{2}{\milli\m}-thick foam layer and has six identical I-shaped slots repeated in the \(z\)-direction (along the static field \(B_0\) in MRI). 
The slots in this microstrip line are required to radiate into the phantom and their length \(L_\text{s}\) affects both the phase constant \(\beta \) and the leakage constant \(\alpha \).
As with any LWA the proposed leaky-wave coil (LWC) delivers some residual power to the end of the transmission line, which has to be absorbed by the matched load. 
To isolate the slotted ground plane of the coil from the phantom, a \SIadj{2}{\centi\m}-thick polycarbonate spacer was used as in a fractionated dipole~\cite{Alex_MRM} designed for the same application and used here for comparison.

\subsection{Design and optimization of the LWC}

\begin{figure*}[tbp]
  \begin{subfigure}[b]{0.32\textwidth}
      \setcounter{subfigure}{0}
      \includegraphics[width=\textwidth]{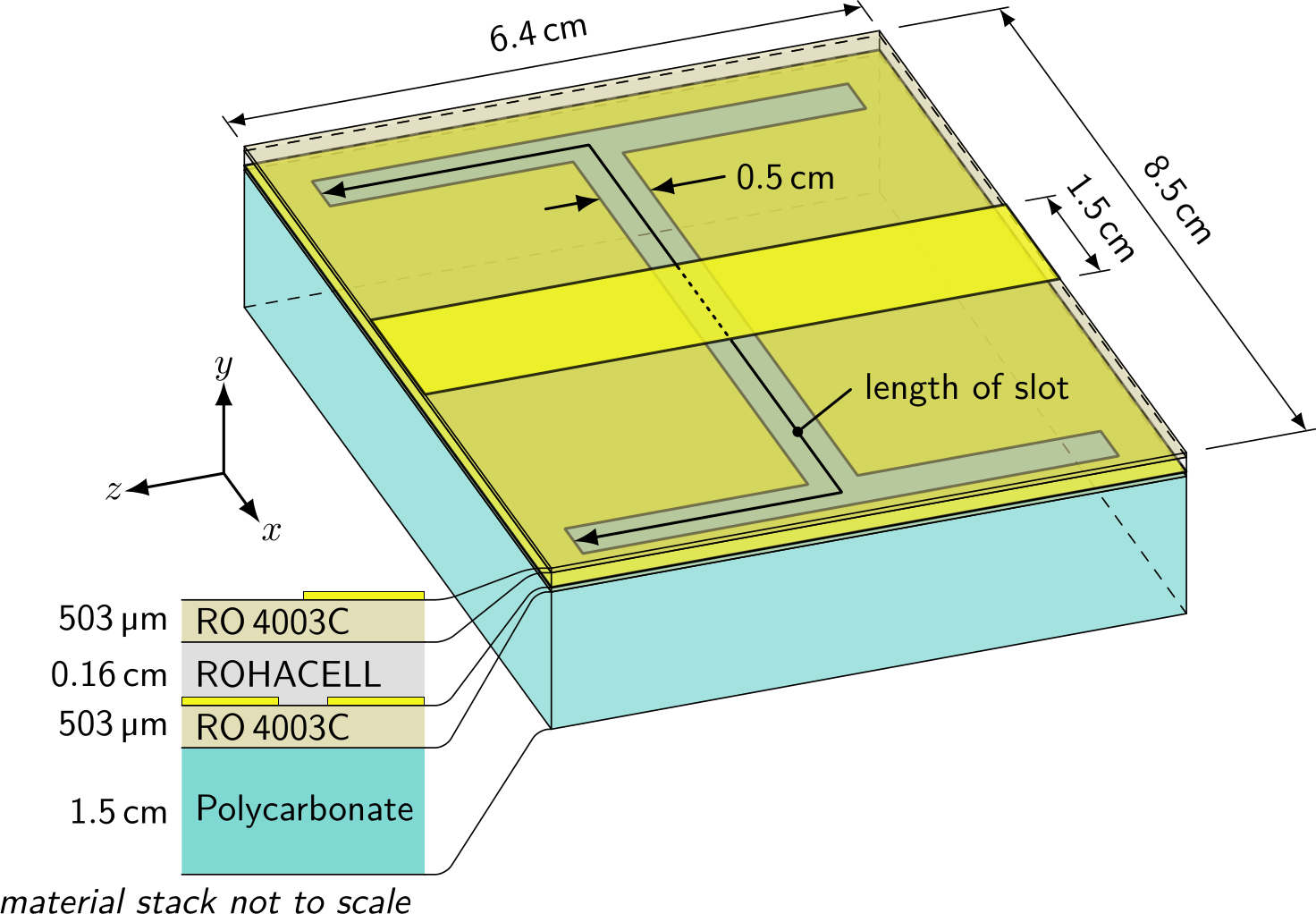}
      \caption{\large Setup of one unit cell}
  \end{subfigure}\hfill
  \begin{subfigure}[b]{0.32\textwidth}
      \setcounter{subfigure}{2}
      \includegraphics[width=\textwidth]{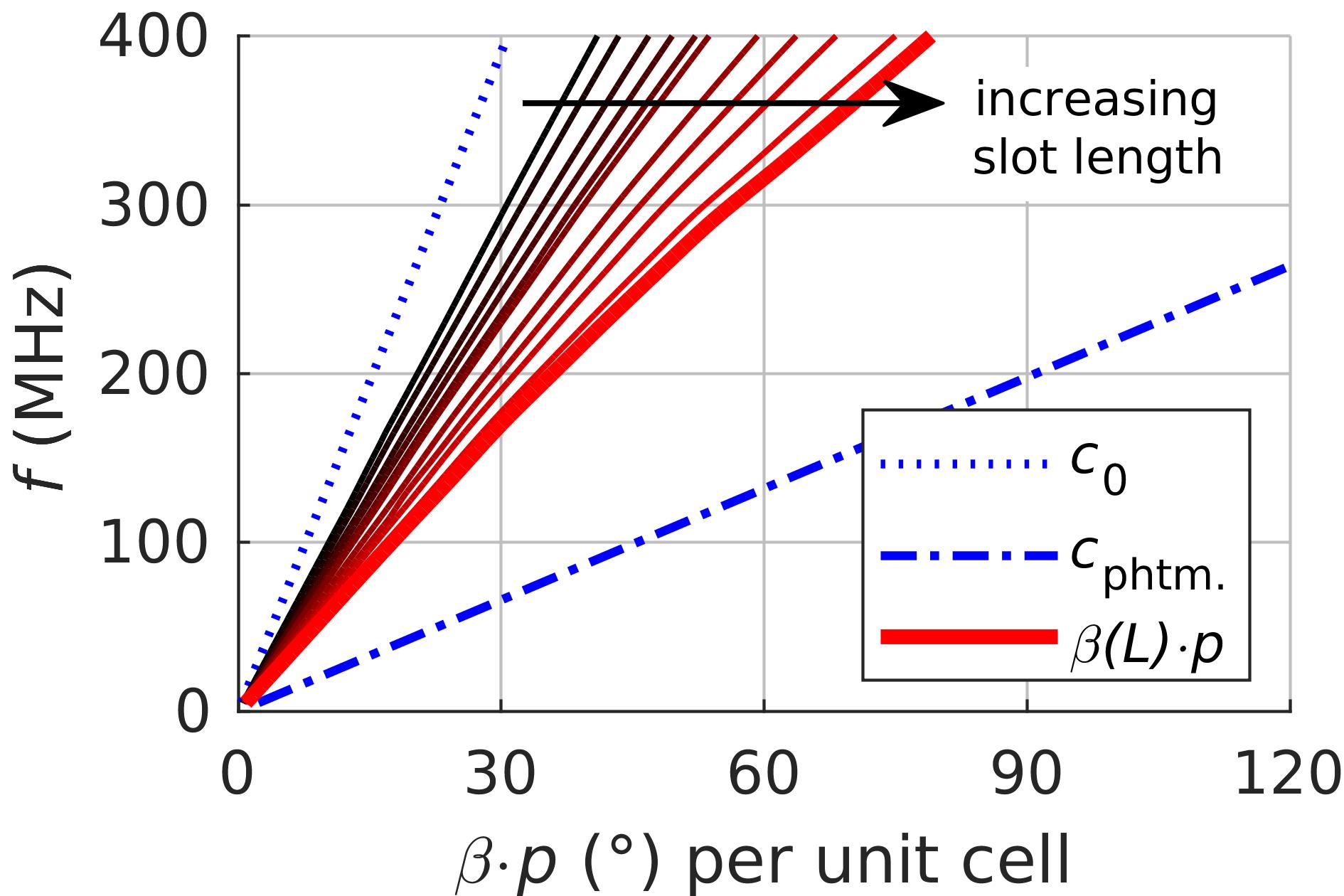}
      \caption{Phase shift; with phantom}
  \end{subfigure}
  \begin{subfigure}[b]{0.32\textwidth}
      \setcounter{subfigure}{4}
      \includegraphics[width=\textwidth]{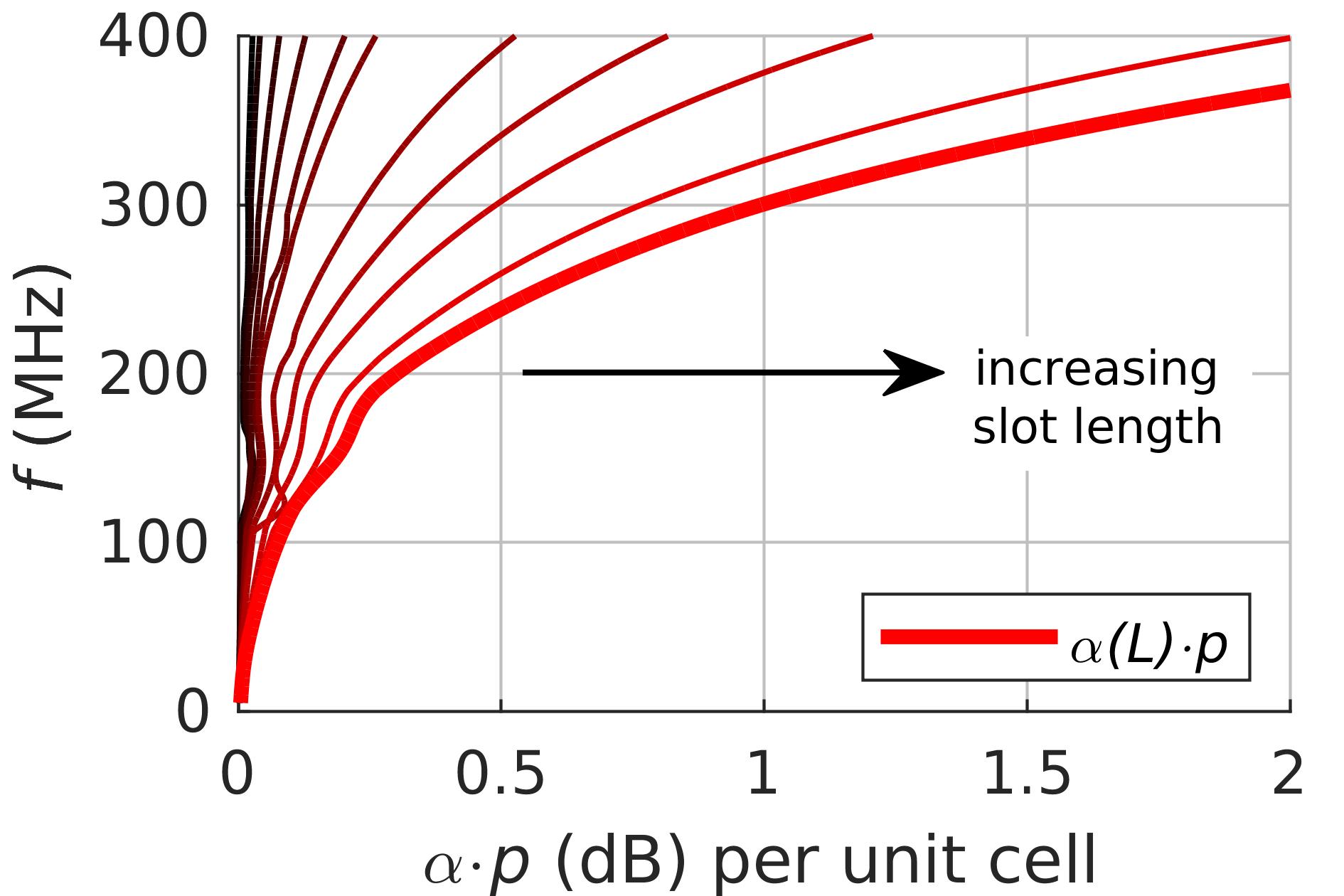}
      \caption{Attenuation; with phantom}
  \end{subfigure}
  \vspace{5mm}
  
  \begin{subfigure}[b]{0.32\textwidth}
      \setcounter{subfigure}{1}
      \includegraphics[width=\textwidth]{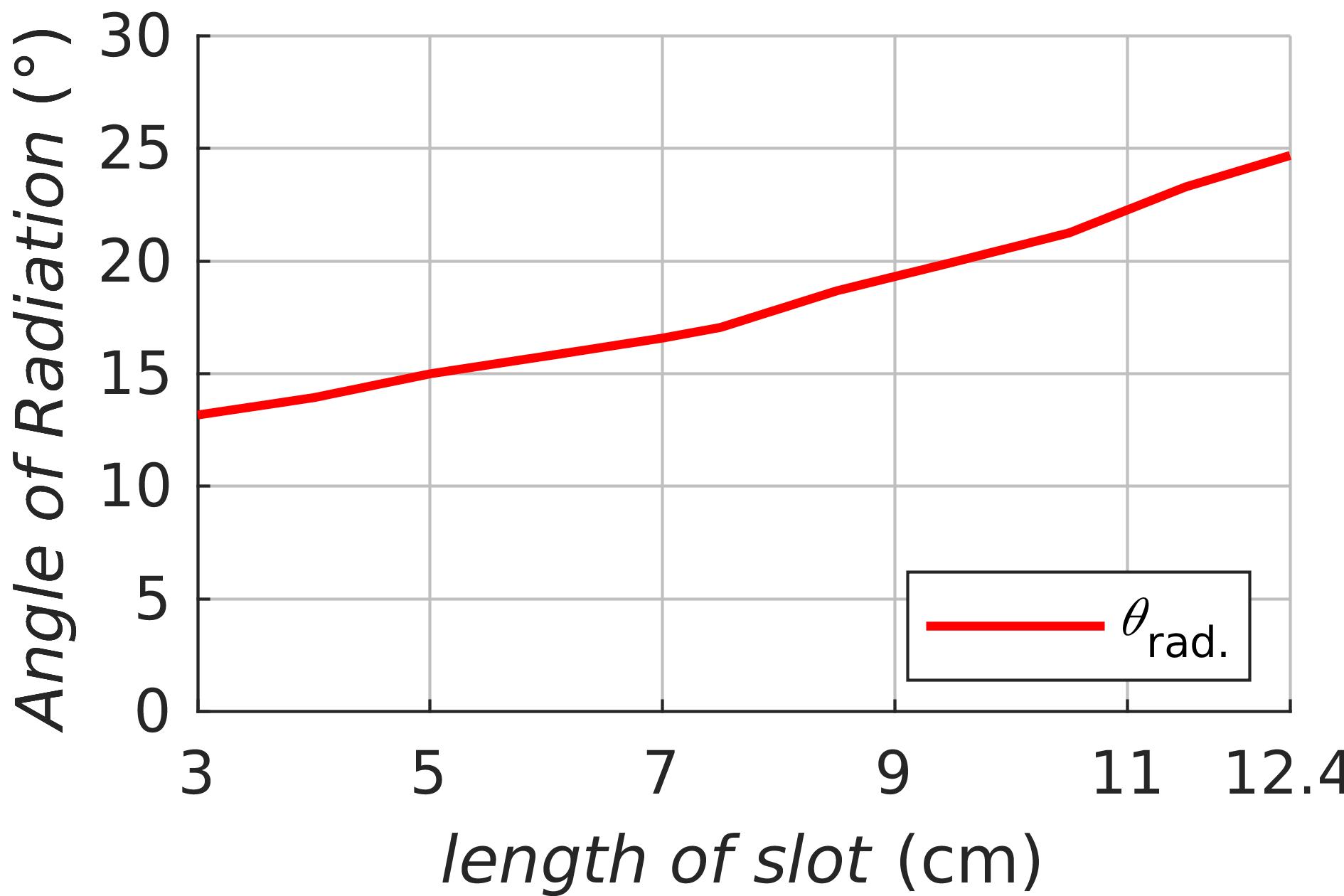}
      \caption{Angle of radiation $\Theta_{\text{rad}}$ vs. $L_{\text{s}}$}
  \end{subfigure}\hfill
  \begin{subfigure}[b]{0.32\textwidth}
      \setcounter{subfigure}{3}
      \includegraphics[width=\textwidth]{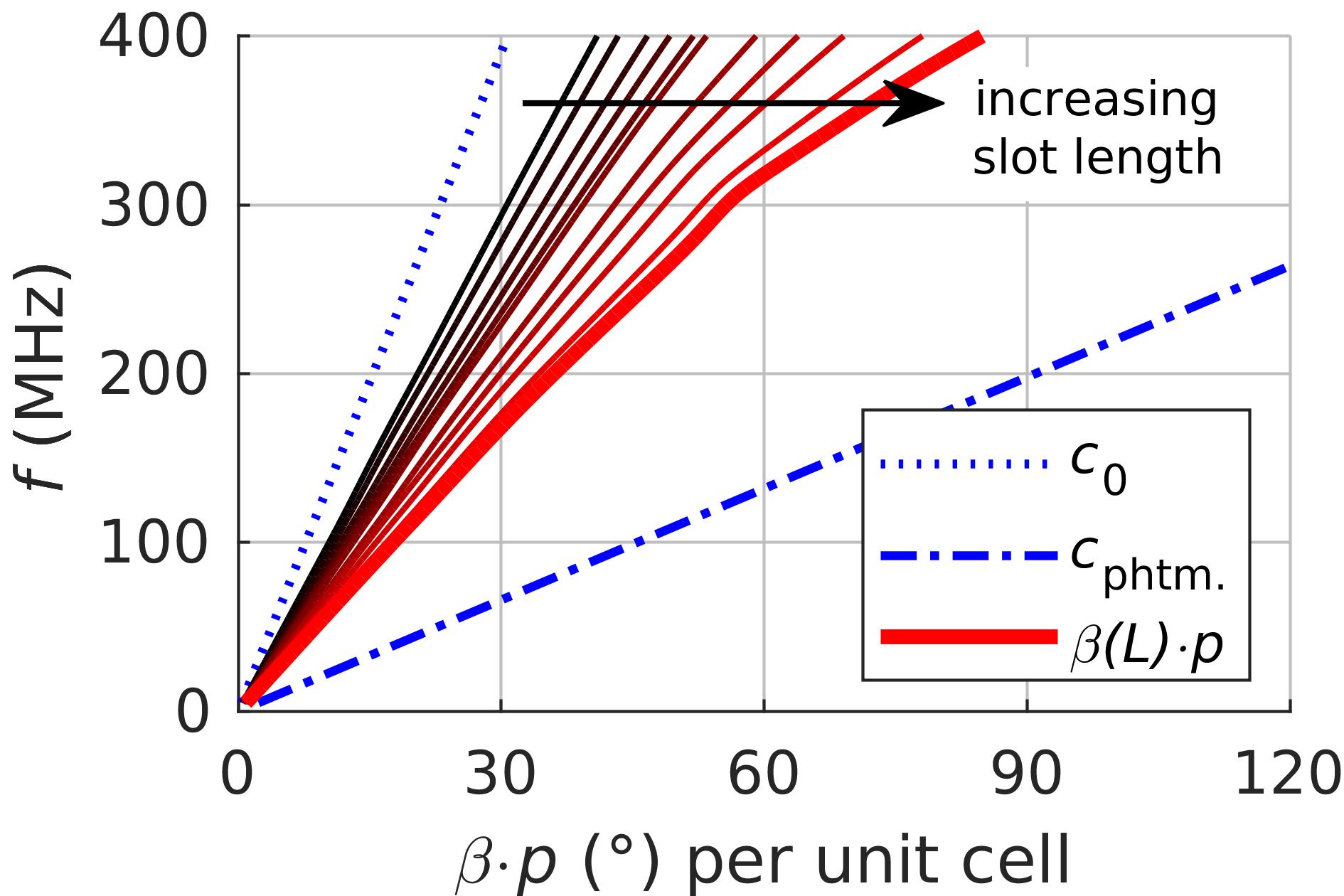}
      \caption{Phase shift; without phantom}
  \end{subfigure}
  \begin{subfigure}[b]{0.32\textwidth}
      \setcounter{subfigure}{5}
      \includegraphics[width=\textwidth]{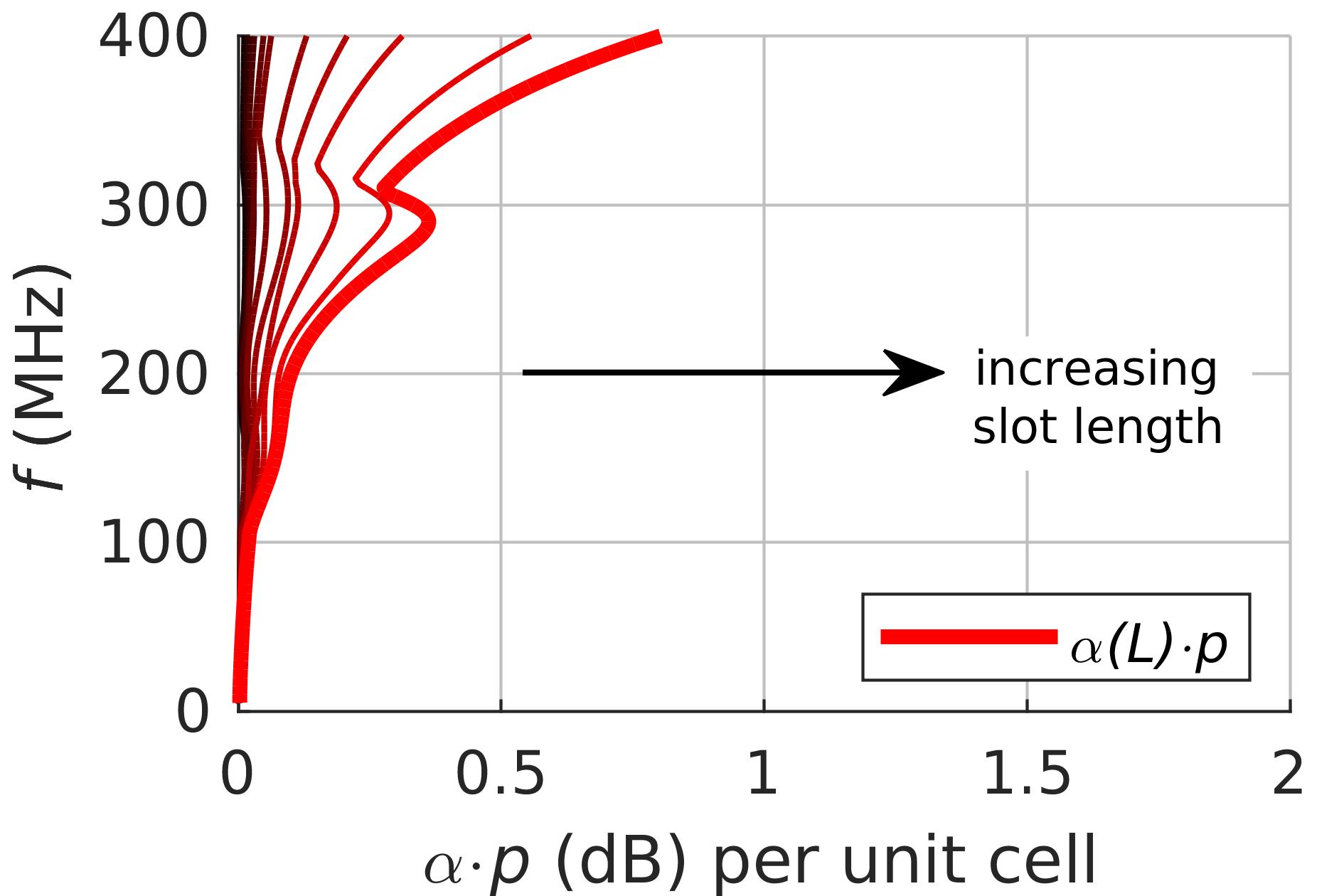}
      \caption{Attenuation; without phantom}
  \end{subfigure}
  \caption{%
    A parametric numerical investigation of dispersion characteristics based on single unit cell (geometry of the setup is shown in (a)): phase and attenuation in the microstrip leaky-wave TL per unit cell (c-f).
    The length of the slot is varied from \SI{3}{\centi\m} to \SI{12.4}{\centi\m}.
    In cases (c) and (e), the TL is loaded with a phantom while cases (d) and (f) represent the unloaded TL.
    The angle of leaky-wave radiation \(\theta_\text{rad}\) in the case of the loaded TL is shown in (b).
  }%
  \label{fig:Dispersion}
\end{figure*}%
The overall dimensions of the LWC were chosen to fit an available space so as to place eight such coils in a pTx 8-element configuration around a human body. 
At the same time the limit to the length for proper positioning on a body was \SI{40}{\centi\m}. 
To obtain a sufficient leakage factor \(\alpha \), the I-shape slots in the ground plane have been arranged with the period of \(p = \SI{6.4}{\centi\m}\) along the microstrip line. 
The length and number of the slots in a single coil were optimized to maximize the \(B_1^{+}\) level at a depth of a prostate for a given transmit power as described in Methods section.  
With this aim, the frequency dispersion of both \(\beta \) and \(\alpha \) in the proposed leaky-wave transmission line were parametrically studied by numerical simulation of a single unit cell depicted in \figref[a]{fig:Dispersion}. 
The results depending on the slots \(L_\text{s}\) are shown in \figref{fig:Dispersion}. 
From the phase constant dispersion curves given in \figref[c]{fig:Dispersion} in the presence of the phantom and in \figref[d]{fig:Dispersion} without the phantom, it is seen that increasing \(L_\text{s}\) slows down the propagating wave. 
The dispersion curve in this case remains below the light line for free space with the phase velocity of \(\mathrm{c}_0\), so that leaky-wave radiation from the line into free space is forbidden. 
The same wave in the microstrip, however, is a fast wave with respect to wave propagation in the phantom medium as all the dispersion curves remain above the corresponding light line with the phase velocity of \(c_\text{phtm}\), which means the leaky-wave radiation to the phantom medium is possible. 
Note that the presence of the phantom has almost no effect on the phase constant \(\beta \) (compare \figrefp[c]{fig:Dispersion} and~\figref*[d]{fig:Dispersion}). 
Conversely, as follows from the comparison of \figrefp[e]{fig:Dispersion} and~\figref*[f]{fig:Dispersion}, the leakage factor \(\alpha \) strongly increases in the presence of the phantom due to the possibility of leaky-wave radiation. 
It can be seen that \(\alpha \) can be continuously adjusted by choosing \(L_\text{s}\). 
Having \(\alpha \) below a particular limit results in lower power efficiency due to dissipation losses in the end load, which decreases the signal level in the ROI. 
Having \(\alpha \) above another limit leads to a very inhomogeneous field pattern in the lateral direction in the phantom.

The slot length \(L_\text{s} = \SI{9.5}{\centi\m}\) was chosen to obtain the attenuation per unit cell \(\alpha \cdot p \approx \SI{-1}{\decibel}\).
In this case, the coil configuration with \(N = \num{6}\) slots with the overall length of \(N \cdot p \approx \SI{38.4}{\centi\m}\) makes the residual power at the end of the line as small as \SI{-6}{\decibel} of the input power.
The corresponding phase constant of the optimized coil \(\beta \approx \SI{15.21}{\per\m}\) leads to a theoretical radiation angle of \(\theta_\text{rad} \approx \SI{24.7}{\degree}\).
This optimized configuration makes the LWC best suitable for prostate imaging in terms of power efficiency.
In the next subsection, the optimized LWC is compared numerically and experimentally to a state-of-the-art coil for prostate imaging at \SI{7}{\tesla}.

\subsection{Leaky-wave coil vs.\ resonant dipole in prostate imaging}

Prostate imaging is one of the most challenging tasks in body MRI at \SI{7}{\tesla}. 
The main challenge in designing surface coils for this application is to maximize the magnitude of \(B_1^{+}\) in the prostate region, which is at a depth of approximately \SI{8}{\centi\m} (comparable to a wavelength), while keeping the peak local SAR close to a body surface as low as possible. 
For comparison, we took a fractionated dipole of length \SI{30}{\centi\m} (see inset in \figref[d]{fig:SAR}), which is the state-of-the-art radiative and resonant coil array element for the application.

\begin{figure}
  \centering
  \includegraphics[width=0.375\linewidth]{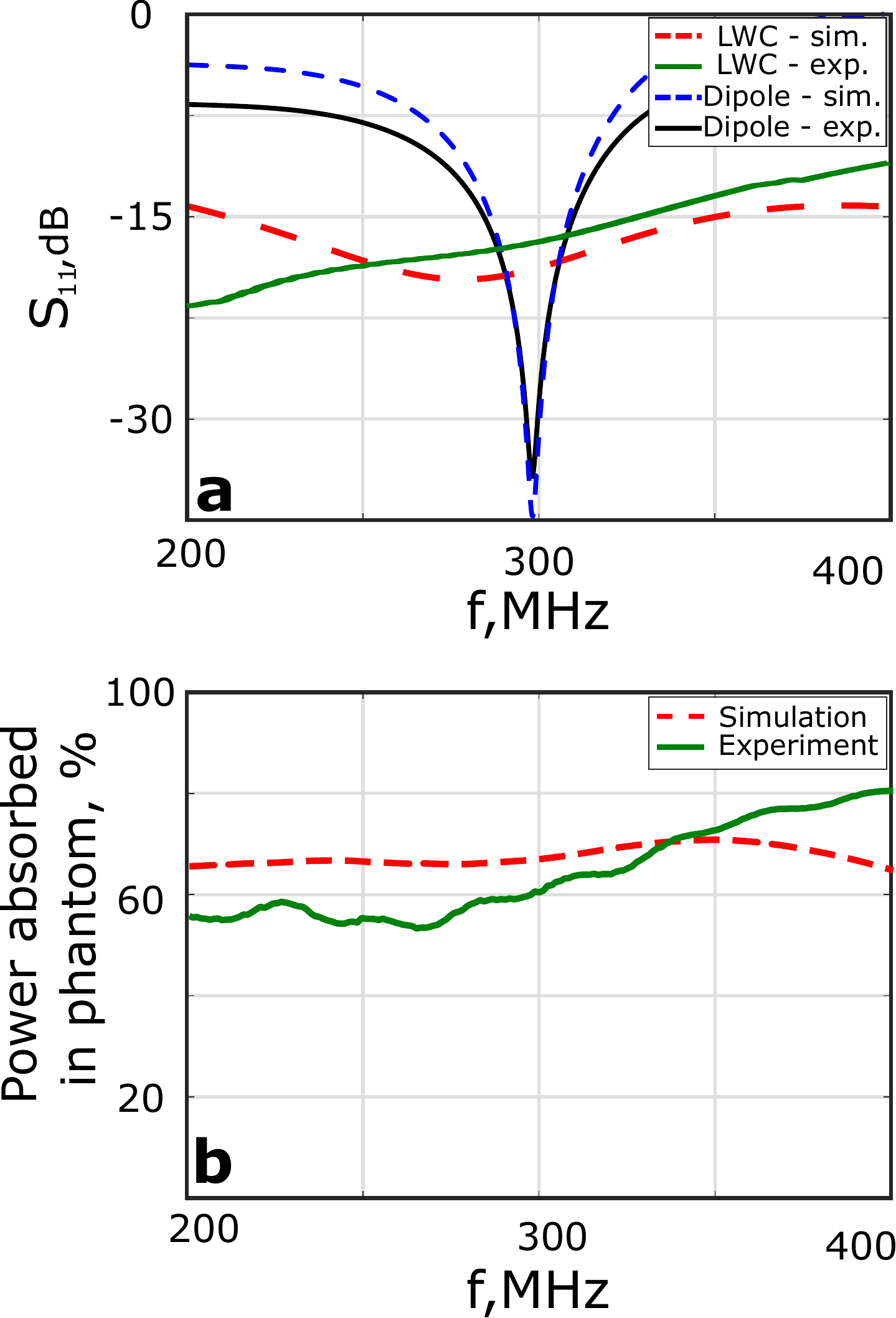}
  \caption{%
    Simulated and measured properties of the optimized LWC with six slots and the reference fractionated dipole: (a) \(\abs{S_{11}}\) at the feed port and (b) relative power \(\nu \) radiated into and absorbed by the subject.
    Simulations were made in the presence of a phantom.
    Measurements were made in the presence of the body of a volunteer.
  }%
  \label{fig:S_param}
\end{figure}%
In the simulations of the LWC placed over the phantom, there were two \SI{50}{\ohm} ports connected to the microstrip line at its beginning (feed) and at the end (matched load). 
The calculated reflection coefficient \(S_{11}\) at a feed port is shown in \figref[a]{fig:S_param} for both coils. 
For the LWC, it is below \SI{-12}{\decibel} in the whole frequency range \SIrange{200}{400}{\mega\Hz}, while for the dipole it remains below \SI{-12}{\decibel} only in \SI{30}{\mega\Hz} band around the central frequency of \SI{300}{\mega\Hz}. 
The relative amount of power radiated into the phantom and absorbed by it calculated as \(\nu = 1- \abs{S_{11}}^2 - \abs{S_{12}}^2\) and shown in \figref[b]{fig:S_param} is above \SI{60}{\percent} and demonstrates the expected non-resonant behaviour of the LWC.
This value is close to one predicted at the design stage based on the selected \(\alpha \). 
Note that the calculated intrinsic dissipation losses in the coil are only \SI{7}{\percent}.

\begin{figure*}
  \centering
  \begin{minipage}{1\linewidth}
  \center{\includegraphics[width=1\linewidth]{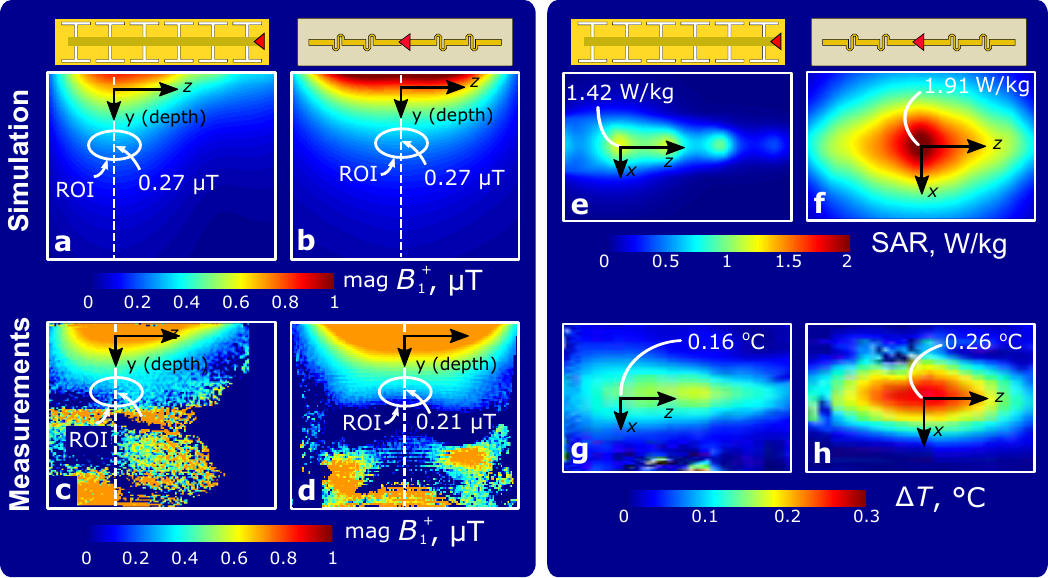}} 
  \end{minipage}
  \caption{%
    Simulated and measured field distributions for \SI{1}{\W} of accepted input transmit power: \(B_{1}^{+}\) patterns in the \(YZ\)-plane of the phantom for the LWC (a,c) and fractionated dipole (b,d); local SAR patterns in the top plane (\(XZ\)) for LWC (e,g) and dipole (f,h).
  }%
  \label{fig:B1_SAR_phantom}
\end{figure*}%
The calculated \(B_{1}^{+}\) distributions in the \(YZ\)-plane of the phantom (see inset in \figref{fig:B1_SAR_phantom}) for the leaky-wave and dipole coils are presented in \figrefp[a]{fig:B1_SAR_phantom} and~\figref*[b]{fig:B1_SAR_phantom}, correspondingly. 
Both coils were simulated with an accepted input power of \SI{1}{\W} and provide almost the same \(B_{1}^{+}\) signal level of \SI{0.27}{\micro\tesla} in the ROI at the depth corresponding to the prostate location in a human body (indicated by the ellipses in \figref{fig:B1_SAR_phantom}).
It can be verified that the two coils have different positions of maximum \(B_{1}^{+}\) with respect to their centers.
While the pattern formed by the dipole is symmetric, the LWC forms an asymmetric pattern with the maximum near to the end load of the microstrip line.

\begin{figure*}[tbp]
  \centering
  \includegraphics[width=0.6\linewidth]{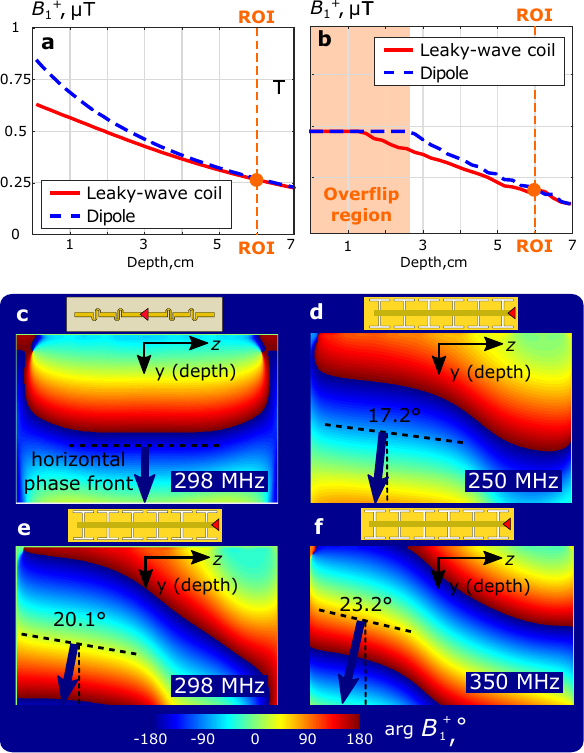}
  \caption{%
    Simulated (a) and measured (b) \(B_{1}^{+}\) vs.\ depth profiles for the LWC and fractionated dipole in the phantom (profiles were plotted along the corresponding dashed lines shown in \figrefp[a]{fig:B1_SAR_phantom} to~\figref*[d]{fig:B1_SAR_phantom}); phase patterns for \(B_{1}^{+}\) numerically calculated in the phantom for the dipole coil at \SI{298}{\mega\Hz} (c), and for the LWC at \num{250} (d), \num{298} (e) and \SI{350}{\mega\Hz} (f).
    The tangent to the phase front for each phase pattern is indicated with a black dashed line.
  }%
  \label{fig:Profiles_Phases}
\end{figure*}%
The calculated \(B_{1}^{+}\) vs.\ depth profiles taken along dashed lines and going through the centers of the ROIs indicated in \figrefp[a]{fig:B1_SAR_phantom} and~\figref*[b]{fig:B1_SAR_phantom} are compared in \figref[a]{fig:Profiles_Phases}.
In the vicinity of the surface, \(B_{1}^{+}\) of the dipole is higher, but starting from a depth of \SI{5}{\centi\m} (including ROI), both coils create the same signal level. The value $\xi$ representing the contribution of reactive fields is compared for the dipole and LWC in Figure \ref{fig:EnergyDiffernceFourCoils}. It is seen that as compared to the dipole and other coil, LWC creates the lowest reactive power contribution at every depth.

The leaky-wave radiation mechanism is clearly illustrated by the simulated phase patterns of the transmit magnetic field created by our coil in the \(YZ\)-plane of the phantom. Unlike the dipole, which creates horizontal phase fronts (see \figref[c]{fig:Profiles_Phases}), the phase fronts of the LWC produce a frequency-dependent angle with respect to the \(y\)-axis.
This effect is similar to frequency beam steering by LWCs in free space.
The behavior of phase fronts for \num{250}, \num{298} and \SI{350}{\mega\Hz} is shown in \figrefp[d]{fig:Profiles_Phases} to~\figref*[f]{fig:Profiles_Phases}. 
The angle of the wave fronts at \SI{298}{\mega\Hz} near the ROI, graphically obtained from the numerically calculated pattern, was \SI{20.1}{\degree}, while its value calculated using an approximate formula (see Methods) was \SI{19.7}{\degree} (see \figref[b]{fig:Dispersion}).

The simulated SAR at an input power of \SI{1}{\W}, distributed in the top plane of the phantom, on which the coil is placed (\(XZ\)-plane according to \figref[a]{fig:SAR}), is shown in \figref{fig:B1_SAR_phantom} for the LWC and the fractionated dipole.
It is seen that the LWC creates \SI{27}{\percent} lower peak local SAR compared to the dipole for the same input power, and, as previously mentioned, for the same \(B_{1}^{+}\) in the ROI.

\begin{figure}[tbp]
  \centering
  \includegraphics[width=0.5\linewidth]{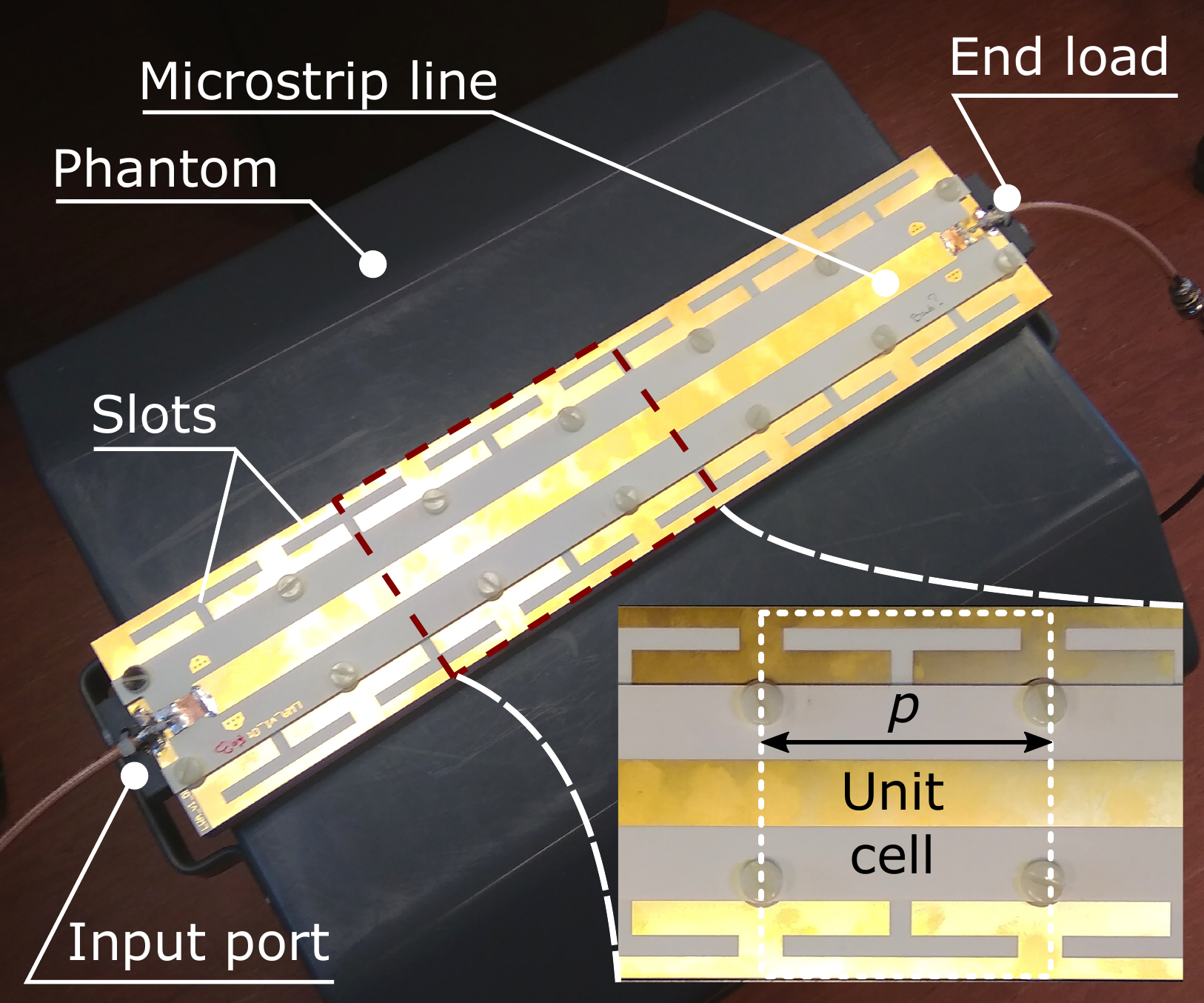}
  \caption{%
    Photo of the fabricated prototype of the optimized LWC placed over a pelvis-shaped homogeneous phantom for MRI characterization.
  }%
  \label{Photo}
\end{figure}%
For the experimental comparison, both the proposed and the reference coils were manufactured.
For the LWC, four identical coils were built to test by \textit{in-vivo} imaging.
One of them is shown in \figref{Photo} placed on the phantom.

The measured \(S_{11}\) and the relative power radiated into a subject, \(\nu = 1-\abs{S_{11}}^2-\abs{S_{12}}^2\) for the proposed coil are compared with corresponding numerically calculated curves in \figrefp[a]{fig:S_param} and~\figref*[b]{fig:S_param}.
In the simulation the coil was placed on the phantom, while in the experiment it was placed on the body of a volunteer.

\(B_{1}^{+}\) patterns measured on a \SI{7}{\tesla} MR system in the same plane as for the previously mentioned numerical results are shown in \figrefp[c]{fig:B1_SAR_phantom} and~\figref*[d]{fig:B1_SAR_phantom}.
As in the simulation, the measured \(B_{1}^{+}\) has an asymmetric pattern for the LWC, and a symmetric one for the dipole. 
At the same time, both coils have the same \(B_{1}^{+}\) field level in the ROI. 
This equivalence of \(B_{1}^{+}\) magnitudes at depths larger than \SI{5}{\milli\m} is additionally confirmed by the measured \(B_{1}^{+}\) vs.\ depth profiles compared for both coils in \figref[b]{fig:Profiles_Phases}. 
In this figure, the orange area (depths smaller than \SI{5}{\centi\m}) corresponds to a saturation of the measured \(B_{1}^{+}\) level where the field levels cannot be compared. 
However for the remaining depths, the measured profiles are very similar to the simulated ones and confirm that both coils have the same penetration of the transmit field.

To support the numerically calculated SAR comparison, temperature increment patterns in the corresponding plane of the phantom (top coronal plane) were  measured by MR thermometry.  Resulting temperature distributions are presented in \figrefp[g]{fig:B1_SAR_phantom} and~\figref*[h]{fig:B1_SAR_phantom}.
For the leaky-wave coil, the temperature increment reached \SI{0.16}{\degreeCelsius} at its maximum, while for the dipole the maximum was \SI{0.26}{\degreeCelsius}, i.\ e.\ \SI{62.5}{\percent} higher for the same input power. 

\begin{figure}
  \centering
  \includegraphics[width=0.6\linewidth]{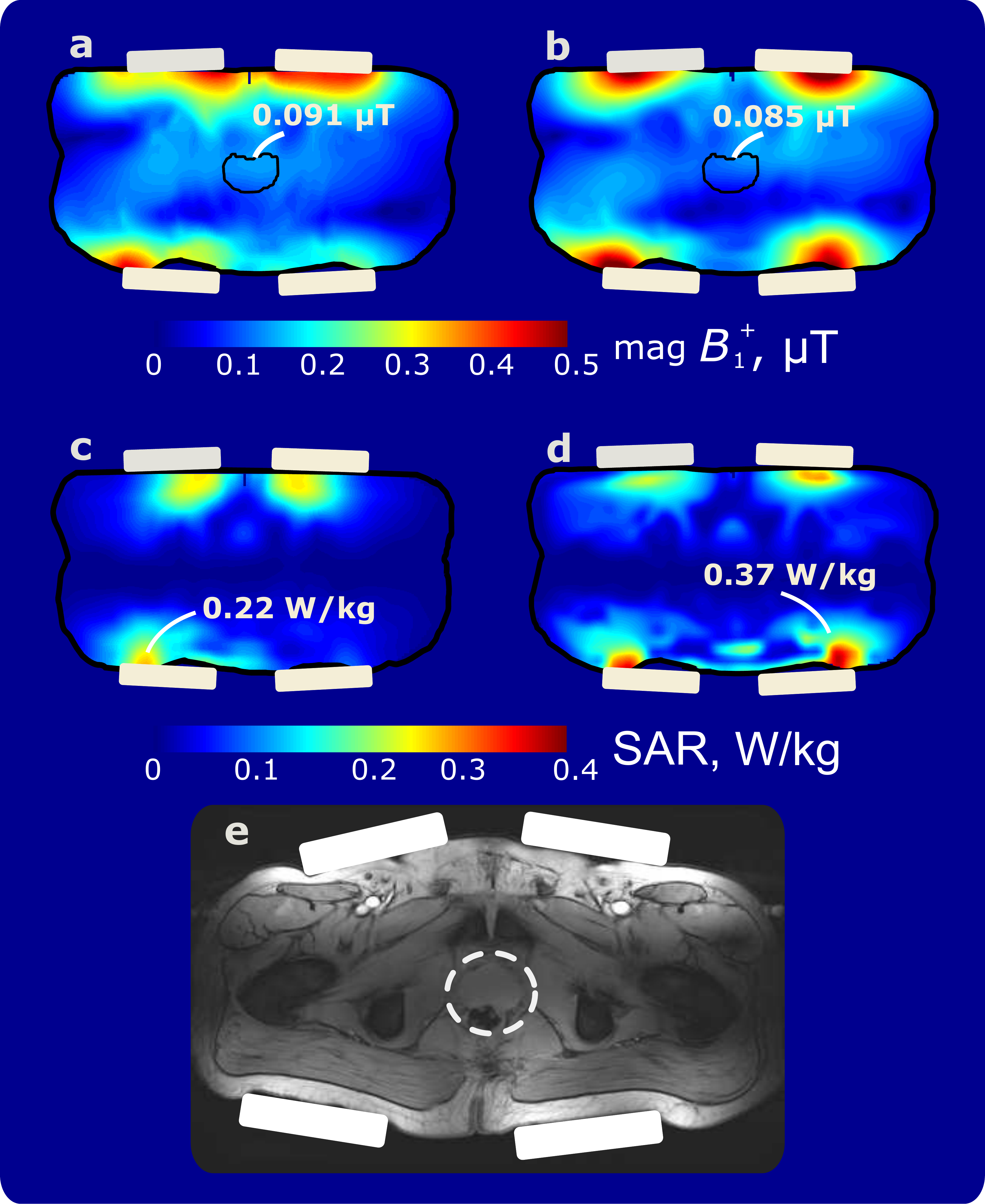}
  \caption{%
    Simulated \(B_{1}^{+}\) for the LWC (a) and fractionated dipole (b) in the transverse slice through the prostate of the human body Duke voxel model for \SI{1}{\W} of accepted power.
    Simulated SAR for the LWC (c) and fractionated dipole (d) in the transverse slice through the maximum of local SAR in the human body Duke voxel model for \SI{1}{\W} of accepted power. (e) \emph{In-vivo} \(T_{1}\)-weighted MR image (transverse slice through the prostate) of a healthy volunteer obtained using a four-element array of LWCs.
  }%
  \label{fig:In_vivo}
\end{figure}%

For the \emph{in-vivo} evaluation of a four-element array, local SAR and \(B_{1}^{+}\) distributions in a multi-tissue human body Duke voxel model were simulated at an input power of \SI{1}{\W}.
A transverse slice of the voxel model through the prostate is shown in \figref{fig:In_vivo} for the LWC (a) and dipole (b).
The corresponding SAR patterns are depicted in \figref{fig:In_vivo} for the LWC (c) and dipole (d).
Simple phase shimming was applied to maximize \(B_{1}^{+}\) in the prostate.
It is seen that the LWC creates \SI{41}{\percent} lower peak local SAR compared to the dipole for the same input power, and \SI{7}{\percent} higher \(B_{1}^{+}\) for the same amount of simulated power.

To demonstrate the performance of the proposed coil in the application, an array of four identical elements used in transceive mode for prostate imaging of a healthy volunteer.
The leaky-wave coils were tightly placed around the body of the volunteer at two locations on the back and two on the stomach.
The obtained \(T_{1}\)-weighted MR image in the transverse plane going through the center of the prostate is shown in \figref[e]{fig:In_vivo}.

\clearpage
\section{Discussion and Conclusion}

In this study, we numerically and experimentally demonstrate the direct correspondence between the resonant properties of surface body coils for MRI with the peak local SAR that they create at the surface of an investigated subject. 
Firstly this was shown by calculating and analyzing the RF-fields created by several coil types, which are the most popular for body imaging at \SI{7}{\tesla}, i.\ e.\ a stripline resonator, a fractionated dipole, a slot and a loop. 
The comparison results of \figref{fig:SAR} clearly show that the broader the impedance matching bandwidth (i.\ e.\ the lower the Q-factor), the lower the peak local SAR at the surface of a homogeneous body phantom for the same level of \(B_{1}^{+}\) at a given depth. 
Among the compared coils the fractionated dipole was the best one in terms of the ratio of \(B_{1}^{+}\) at the depth of a prostate to the square root of the peak local SAR (at the surface hotspot). 
The higher peak SAR of narrower-band coils can be explained by the effect of quasistatic RF-fields in the near field region. 
It is well-known from antenna theory that the matching bandwidth is associated with reactive fields storing some energy of electric and magnetic fields in the vicinity of an antenna. 
At the antenna's resonance, the electric and magnetic energies are equal, which can result in a real input impedance at the resonant frequency. 
However, the amount of the electric and magnetic energy itself strongly depends on the antenna geometry and is related to the bandwidth in which the antenna can be matched to its source.

In the case of UHF-MRI of the prostate, the ROI is located at a depth comparable to the wavelength in body tissues. 
In other words, it is located in the intermediate-to-farfield region. 
At the same time, the peak local SAR hotspot is typically located near the surface of the subject, which is inside the nearfield region. 
Human body tissues represent a highly-conductive medium, which has a complex propagation constant and characteristic impedance. 
As a result, RF-fields cannot be easily decomposed into quasi-static and radiation components as in free space. 
However, as one can see from \figref{fig:SAR}, the weak electric field in the near region of the most broadband coil, a dipole, causes the lowest SAR and the highest radiative magnetic field at large depths among the coils tested. 
Moreover, from Figure \ref{fig:EnergyDiffernceFourCoils} it is evident that the dipole coil has the lowest value of $\xi$ at any depth, which is due to the lowest contribution of reactive, spatially separated electric and magnetic fields.
This leads to the conclusion that a reduction of the peak local SAR, which is the main limiting factor for any transmit coils for UHF body imaging, requires the quasi-static electric field to be minimized. 
As in antennas for free space, for surface body coils, lower quasi-static fields should mean lower Q-factors. 
We have additionally proven this conclusion by introducing and demonstrating the first non-resonant surface coil for body imaging, the \emph{leaky-wave coil}.

Our LWC is based on the continuous energy leakage mechanism and employs no standing waves. 
By propagating from one end of a microstrip to the other, most of the power leaks into the subject and is absorbed within. 
The discrete slots of the ground plane being excited by the propagating wave radiate power to the subject and behave as a slot array with almost linear phase spatial variation. 
As a result, a beam of the field at large depths is produced at an angle which changes with frequency, as seen from the simulations in \figrefp[d]{fig:Profiles_Phases} to~\figref*[f]{fig:Profiles_Phases}. 
The existence of this non-resonant radiation mechanism was demonstrated experimentally with very good agreement between simulated and measured field patterns and S-parameters of the LWC.

If the microstrip line is matched to the input port and loaded to a matched impedance at the end, the coil operates as a continuous junction between the transmitter and the medium of human body tissues. 
The real part of the characteristic impedance of the LWC stays close to \SI{50}{\ohm} and varies slowly with frequency, which additionally shows the non-resonant properties.

The study of the transmit signal field \(B_{1}^{+}\) in a homogeneous phantom shows that the proposed LWC creates the same level of magnetic field at the depth of a prostate. 
This result obtained from simulations on a phantom was precisely confirmed by measurements. 
However, as shown by temperature increment measurements in the same phantom, the LWC causes just above half of the heating in the near field region than the dipole, which is in good agreement with local SAR simulations. 
This means that the LWC is more efficient as a transmit coil than the dipole due to lower quasi-static fields. 
In fact, in the vicinity of the subject surface, both the electric and magnetic fields are reduced for the proposed coil, as was additionally supported by the results of Figure \ref{fig:EnergyDiffernceFourCoils}.
However, for a depth larger than \SI{5}{\centi\m} (like the prostate location in the human body), the \(B_{1}^{+}\) of both coils at a given transmit power is the same. 
The advantage in SAR performance has been verified by running a simulation of a four-element array of LWCs placed on a detailed voxel model of a human body. 
The same simulation was done using fractionated dipoles. 
In contrast to a homogeneous phantom, the body model has very inhomogeneous distributions of conductivity and permittivity. 
As a result, the positions of the SAR hotspot change. 
But still, the LWC array induces  \SI{41}{\percent} lower peak local SAR as can be seen in \figref[c]{fig:In_vivo}.

Apart from reduced local reactive fields and smaller peak SAR, the proposed radiation mechanism has the advantage of broadband and very stable impedance matching. 
The arrangement of non-resonant slots couples the wave propagating in the microstrip line to the wave propagating in the medium of body tissues. 
For the phantom used in the simulations and measurements, the wave impedance of a propagating wave was complex~\cite{king1960half} and equal to \(\eta_\text{ph} = \SI{58+j19}{\ohm}\) at \SI{298}{\mega\Hz}. 
Our results showed that the LWC was effectively loaded by this wave impedance and gave a very stable input impedance close to \SI{50}{\ohm} at the input port. 
As a result, the coils provided a return loss of less than \SI{-12}{\decibel} in the entire frequency band.
Additional simulations have shown that this coil impedance depends mostly on the permittivity of the subject and the best intrinsic matching is obtained when the real part of the wave impedance of the phantom material is close to \SI{50}{\ohm}. 
Conveniently, the average wave impedance of human body tissues is indeed close to \SI{50}{\ohm}, which enables the self-matching capability of our LWC.
To our knowledge, the leaky-wave coil is the only surface coil which does not require any matching network. 
Furthermore, the input impedance of the LWC in the experiment was stable with respect to any variation of the subject's properties. 
There was no significant change in the port's input impedance value when the phantom was replaced by the body of a volunteer as shown in \figref{fig:S_param}.

From an engineering point of view, another advantage of the leaky-wave coil is that it operates using a radiative microstrip transmission line with an asymmetric topology. 
This means that it does not require a balun, i.e. device that converts non-symmetric to symmetric signals, at its input when fed by a coaxial cable soldered to the microstrip.

In summary, we proposed and demonstrated a non-resonant surface coil based on the leaky-wave radiation mechanism. 
It was clearly shown that due to lower near electric fields, it reduces peak SAR to just over a half for the same transmit field in the ROI. 
Moreover, the new coil does not require either a matching network or a balun and is very stable with respect to variation in the subject. 
This allows us to conclude that non-resonant surface coils for UHF MRI, in particular leaky-wave coils, are very attractive for future clinical applications of UHF MRI.

\section{Methods}

\subsection{Optimization and simulation}

For the numerical analysis and optimization of unit cells (slots in the microstrip line) the FDTD-based solver EMPIRE-XPU-2018 (IMST, Kamp-Lintfort, Germany) was utilized.
The propagation characteristics were calculated by extracted ABCD-parameters of one unit cell (see \figref{fig:Dispersion}) embedded in a fitting microstrip line, i.\ e.\ a continuous microstrip line of the same cross-section, but without slots, where the port reference planes are positioned at the terminals of the unit cell~\cite{Caloz.Itoh_ElectromagneticMetamaterialsTransmission_2006,Svejda_DualfrequenteCRLHMetaleitungsResonatorenfur_2019}.
The optimal unit cell balances the residual power loss at the end of the line and the field pattern.
Short slots lead to weak radiation and high losses at the end, i.\ e.\ to small \(\alpha \).
The large length \(L_\text{s}\) of the slots makes them close to their self-resonance, which increases the leakage constant, but it also changes the angle of radiation  (see \figref{fig:Dispersion}) by changing the phase constant \(\beta \).
Large angles \(\theta_\text{rad}\) with respect to the \(y\)-axis lead to distortions of the magnetic field pattern and lower \(B_{1}^{+}\) for the same input power.
As a result, based on the goal to maximize \(B_{1}^{+}\) in the ROI, an appropriate value of \(L_\text{s}\) and \(\alpha \) are first selected from the calculated family of dispersion curves (see \figref[e]{fig:Dispersion}).
It is worth noting that the I-shaped slots were chosen to fit to the maximum possible width of the coil (\SI{8.5}{\centi\m}), which allows the accommodation of 8 such coils around a human body to be able to use an 8-element pTx array configuration.

Next, the corresponding \(\beta \) can be found from the curve family for \(\beta \) in Fig.~\ref{fig:Dispersion}(c) to estimate the radiation angle.
The angle \(\theta_\text{rad}\) between the \(y\)-axis and the propagation direction in the phantom is estimated using the following formula~\cite{Caloz.Itoh_ElectromagneticMetamaterialsTransmission_2006}:
\begin{equation}
  \theta_\text{rad} = \arcsin\left({\frac{\beta}{k_0 \sqrt{\varepsilon_\text{r}}}}\right)
\end{equation}
This formula is approximate, as the wave vector in the conductive medium of the phantom is actually complex, so one should use both the propagation and attenuation vectors to be matched in the medium and the transmission line~\cite{baccarelli2018analytical}.

Simulations of the proposed coil and the reference dipole coil in the presence of a phantom and a human body model were performed in CST Microwave Studio 2019 (CST, Darmstadt, Germany).
For simulations with a phantom, adaptive tetrahedral meshing was performed at the Larmor frequency of \SI{298}{\mega\Hz} using the finite-element method (Frequency Domain Solver). 
The number of meshcells was approximately \num{700000} for all simulated models.

For the safety assessment, four element arrays of leaky-wave and dipole coils were simulated with the same software on the human body Duke voxel model (ITIS Foundation, Zurich, Switzerland) with the finite integration method (Time Domain Solver) in CST Microwave Studio 2019.
The number of meshcells was approximately \num{30e6} for all simulated models.

\subsection{Coil prototyping and on-bench measurements}

The prototype of the leaky-wave coil was composed of two Rogers RO4003 (\(\varepsilon_\text{r} = \num{3.38}\), \(\tan{\delta} = \num{0.003}\)) \SIadj{0.203}{\milli\m}-thick single-layer PCBs, separated by a \SIadj{2}{\milli\m}-thick foam layer of ROHACELL\,31HF.
The top PCB was used to form the line strip, while the bottom one formed the ground plane with 6 I-shaped etched slots.
Due to the layer of \SIadj{20}{\milli\m}-thick polycarbonate below the ground, the self-resonance of the slots provided a stable phase constant \(\beta \) in the presence and absence of the phantom.
Note that the same polycarbonate spacer was used in the reference dipole coil.

The LWC was fed from one side with a \SI{50}{\ohm} coaxial cable soldered to the microstrip.
The second port of the LWC was loaded by a \SI{50}{\ohm} high power resistor.
In the prototype neither a matching network nor a balun was used.
On the contrary, the dipole required a lumped-element network to be matched to \SI{50}{\ohm} and a ceramic balun placed on the feeding coax.

For all measurements a pelvis-shaped homogeneous phantom with similar sizes and electrical properties as in numerical simulations was used.
S-parameters measurements of the leaky-wave and dipole coils were measured using a Copper Mountain TR1300/1 2-Port vector network analyzer.

\subsection{MRI characterization}

Measurements of \(B_{1}^{+}\) and the temperature increment in a phantom as well as MR imaging of a healthy volunteer were performed on a \SI{7}{\tesla} Philips Achieva  MRI-platform (Philips Healthcare, Best, The Netherlands) at the University Medical Center Utrecht, the Netherlands. 

First, a homogeneous phantom was used to measure \(B_{1}^{+}\) patterns (\figrefp[a]{fig:B1_SAR_phantom},~\figref*[b]{fig:B1_SAR_phantom} and~\figref*[e]{fig:B1_SAR_phantom}) created by the single leaky-wave and dipole coils located on the top of the phantom.
The \(B_{1}\) maps were created using the Dual-TR method~\cite{Double_TR} (AFI) with the following scan parameters: Field of View: \SI{231 x 400 x 140}{\milli\m}, voxel size: \SI{2.2 x 3.8 x 10}{\milli\m}.
TE/TR1/TR2: \SIlist{2.2; 50; 250}{\micro\s}. 

The temperature maps were created using the proton resonance frequency shift method (\figrefp[c]{fig:B1_SAR_phantom} and~\figref*[d]{fig:B1_SAR_phantom})~\cite{Thermometer}.
The phantom was placed in the scanner room 1 hour in advance to reach thermal equilibrium.
Heating was provided by off-resonance (\(+\SI{100}{\kilo\Hz}\)) block pulses with a power of \SI{108}{\W} on a \SI{10}{\percent} duty cycle, yielding an average power of \SI{10.8}{\W}.
The following scan parameters were used: Field of View: \SI{230 x 348 x 414}{\milli\m}, voxel size: \SI{3.6 x 4.8 x 6}{\milli\m}, TE/TR: \SIlist{10; 15}{\micro\s}.

The \emph{in-vivo} study of a healthy volunteer was approved by the local medical ethics committee and informed consent was obtained from the subject.
\(T_2\)-weighted in-vivo body images (TR/TE = \SIlist{2500; 90}{\milli\s}, \SI{0.5 x 0.5 x 4}{\milli\m}, TSE-factor: 9) were obtained using the manufactured LWC array and fractionated dipole array.

\section{Acknowledgements}
This project received funding from the European Union's Horizon 2020 research and innovation program under grant agreement no.\ 736937. 

The authors thank Bart Steensma, Redha Abdeddaim, Constantin Simovski, Stefano Maci, Chris Collins, Ksenia Lezennikova and the team of the RF-coils lab at UMC Utrecht for their help and useful discussions.

\bibliography{references}

\end{document}